\documentclass[conference]{IEEEtran}
\IEEEoverridecommandlockouts
\usepackage{cite}
\usepackage{amsmath,amssymb,amsfonts}
\usepackage{algorithmic}
\usepackage{graphicx}
\usepackage{textcomp}
\usepackage{xcolor}
\usepackage{url}
\usepackage{microtype}

\def\BibTeX{{\rm B\kern-.05em{\sc i\kern-.025em b}\kern-.08em
    T\kern-.1667em\lower.7ex\hbox{E}\kern-.125emX}}
\begin{document}

\title{On The Effectiveness of the UK NIS Regulations as a Mandatory Cybersecurity Reporting Regime

 \thanks{This research was funded by NCSC, part of GCHQ.}
}

 \author{\IEEEauthorblockN{Junade Ali}
 \IEEEauthorblockA{\textit{The Alan Turing Institute} \\
 \textit{Defence and National Security Programme}\\
 London, United Kingdom \\
 jali@turing.ac.uk}
 \and
 \IEEEauthorblockN{Chris Hicks}
 \IEEEauthorblockA{\textit{The Alan Turing Institute} \\
 \textit{Defence and National Security Programme}\\
 London, United Kingdom \\
 c.hicks@turing.ac.uk}
 }

\maketitle

\begin{abstract}
Existing cybersecurity literature lacks a source of empirical, representative data as to the true nature of cyberattacks on Critical National Infrastructure. We have obtained UK-wide data on incidents reported under the Network and Information Systems (NIS) Regulations in 2024 causing ``a significant impact on the continuity'' of essential services and comparator data from intelligence agencies. We find that 29\% of NIS reports already concern cybersecurity incidents. As the UK Government seeks to extend cybersecurity reporting, we find the NIS Regulations are limited in their effectiveness; whilst our requests revealed 30 cybersecurity incidents reported under the NIS regulations, there were 89 incidents classified as ``highly significant and significant'' captured by the National Cyber Security Centre in the 2024 reporting year. Whereas 36\% of Cybersecurity and Infrastructure Security Agency reported attacks concerned espionage, from NIS data we find 100\% NIS-reportable cyberattacks concerning healthcare systems in England in 2024 were ransomware.
\end{abstract}


\section{Introduction}
Empirical, representative data on the behaviour of attackers remains an enigma in cybersecurity. Existing large-scale datasets often contain vast amounts of media-reported data, which may not be representative of the true picture of cyberattacks on Critical National Infrastructure (CNI)~\cite{national2025cybercrime, sangari2022modeling}. However, for cybersecurity research and policy work to be appropriately targeted, empirical data from real-world targets is of enormous value~\cite{cremer2022cyber}.

NIS Regulations in the UK~\cite{nis_regulations} provide for mandatory reporting of network and information systems incidents which have ``a significant impact on the continuity'' of essential services (including cybersecurity incidents). Under this framework, operators of essential services (OES) and relevant digital service providers (RDSPs) are required to notify a sector-specific designated competent authority (DCA) - for example, the Department of Health and Social Care for healthcare in England, or Ofcom for digital infrastructure - which then forwards reports to GCHQ as the ``single point of contact'' (as an annual report). However, GCHQ is not subject to Freedom of Information (FOI) legislation.

In this paper, we seek to overcome this challenge using two techniques absent from existing literature. Firstly, we obtain data related to incidents reported under the NIS Regulations by subjecting each of the DCAs to FOI requests. Secondly, as a complementary source, we have aggregated data from Cybersecurity and Infrastructure Security Agency (CISA) advisories to identify the initial access vector and attack type information of attacks which the Five Eyes intelligence communities have chosen to disclose.

Using this approach, we seek to answer the following questions:

\begin{itemize}
    \item \textbf{RQ1:} Can mandatory cybersecurity reporting regimes be combined with freedom of information laws to assist researchers in collecting data on cyberattacks?
    \item \textbf{RQ2:} What cybersecurity incidents are captured by mandatory reporting under the UK NIS Regulations during 2024, and to what extent does the resulting data describe attack types and initial access vectors affecting UK CNI?
    \item \textbf{RQ3:} Analysing curated CISA advisories in 2024, what types of incidents have been selected for disclosure?
\end{itemize}

To assist readability, a glossary of key abbreviations is provided in Table~\ref{tab:abbreviations}.

\begin{table}[h]
\caption{Abbreviations}
\label{tab:abbreviations}
\centering
\scriptsize
\setlength{\tabcolsep}{4pt}
\renewcommand{\arraystretch}{1.1}
\begin{tabular}{@{}lp{0.78\linewidth}@{}}
\hline
\textbf{Term} & \textbf{Meaning} \\
\hline
CER & Critical Entities Resilience Directive (EU) \\
CISA & Cybersecurity and Infrastructure Security Agency (US) \\
CNI  & Critical National Infrastructure \\
CTI & Cyber Threat Intelligence  \\
DCA  & Designated Competent Authority (under NIS regulations) \\
DHSC & Department of Health and Social Care (UK) \\
DI   & Digital Infrastructure \\
DoF  & Department of Finance (Northern Ireland) \\
DORA & Digital Operational Resilience Act (EU) \\
DWQR & Drinking Water Quality Regulator (Scotland) \\
ENISA & European Union Agency for Cybersecurity \\
FIMI & Foreign Information Manipulations and Interference \\
FOI  & Freedom of Information \\
FOIA & Freedom of Information Act 2000 (UK) \\
FOISA & Freedom of Information (Scotland) Act 2002 (Scotland) \\
GCHQ & Government Communications Headquarters (UK) \\
MCA  & Maritime and Coastguard Agency (UK) \\
NCSC & National Cyber Security Centre (part of GCHQ) \\
NCCS & Network Code on Cybersecurity (EU) \\
NIS  & Network and Information Systems (EU \& UK Regulations) \\
NIS2 & Network and Information Systems (EU Directive) \\
OES  & Operators of Essential Services (under NIS Regulations) \\
Ofcom & Office of Communications (UK) \\
RDSP & Relevant Digital Service Providers (under NIS Regulations) \\
RVA & Risk and Vulnerability Assessments (by CISA) \\
\hline
\end{tabular}
\end{table}

\section{Related Work}
\subsection{Reporting Regimes \& NIS Regulations}
The NIS Regulations~\cite{nis_regulations} form part of retained European Union (EU) law in the UK. In part, they require ``operators of essential services'' in certain sectors to disclose to the relevant DCA (and in the case of relevant digital services, the Information Commissioner) ``any incident which has a significant impact on the continuity of the essential service which that OES provides''. The thresholds are typically determined in guidance from the DCA. For example; whilst the Information Commissioner sets thresholds tailored to digital services~\cite{ico_reporting}, the Department for Health and Social Care (in relation to England) sets the thresholds based on measures related to healthcare (e.g. $> 0$ excess fatalities/casualties, or $> 50$ patients at risk of potential clinical harm)~\cite{dhsc_reporting}.

In 2025, the Government announced its intention to put a Cyber Security and Resilience Bill before Parliament to extend these requirements, as has been done in the EU with the implementation of the NIS2 Directive~\cite{cybersecurity_policy_statement}.

Further, from 14th January 2025 to 8th April 2025, the UK Government consulted on mandatory reporting of ransomware~\cite{government_ransomware_consultation}, specifically to ``provide the Government with an initial report within 72 hours of the attack, covering key details, and a more in-depth report within 28 days''. 63\% of respondents (both individual and organisational) agreed with the proposal of: ``Economy-wide mandatory reporting for all organisations and individuals''.

Following the UK's departure from the EU, the NIS Cooperation Group published a report of EU-wide NIS incidents~\cite{eu_nis_report} in August 2025, tracking coarse root causes (i.e. 51\% system failures and 37\% malicious actions), however this is not a cybersecurity-specific report and provides no granularity as to the attack types or initial access vector information, noting ``2025 will be covered by NIS2 Directive (NIS2) and incident reporting process is continuously improving and maturing''. The authors note: ``In terms of processes, there are still more synergies to be explored with reporting in other sectors, under other pieces of legislation, such as DORA, the Network Code on Cybersecurity (NCCS) and the CER Directive''.

Accordingly, there is currently no data providing a UK-wide picture of NIS reports specifically related to cybersecurity. This means there is a key gap for policy makers in understanding the effectiveness of existing reporting regimes and for researchers to target research at the most high-reward components of cybersecurity.

\subsection{The Need for Better Data}
Against this foreground, there is a backdrop of work highlighting the importance of data to target cybersecurity research. A 2025 consensus study report from the US National Academies~\cite{national2025cybercrime} has noted: ``comprehensive, consistent, and reliable data and metrics on cybercrime still do not exist - a consequence of a shortage of vital information resulting from the decentralized nature of relevant data collection at the national level''. The work highlights the importance of more datasets of greater diversity in source, arguing: ``an exact enumeration of cybercrime from any single data resource is infeasible''.

A systematic review on data availability of cyber risk and cybersecurity~\cite{cremer2022cyber} posits that: ``the lack of available data on cyber risk poses a serious problem for stakeholders seeking to tackle this issue''.

As will be discussed later, there is a volume of material based on public media reports of cybersecurity incidents, but~\cite{sangari2022modeling} estimates that only $\approx 3\%$ of incidents appear to be publicly reported, biased by revenue, industry and incident type.

A potential cause for this can be found in~\cite{ali2025so} which explores the phenomenon of \textit{loss aversion}, whereby humans will feel the pain of a loss to a greater extent than the pleasure of a gain, leading to issues being covered-up. This is despite the fact that a representative opinion polling of software engineers found the ability to discuss and address problems quickly, when they emerge, was correlated with an $87\%$ higher success rate in software project success. Similar research in cybersecurity has echoed these ideals with~\cite{ebert2025learning} arguing that ``appropriate trust-building mechanisms such as confidentiality, impunity, and independence should be anchored in the system''.

\subsection{Impact of AI}
There has been increasing work on the use of AI techniques such as reinforcement learning~\cite{foley2022autonomous} and AI agent attackers/defenders~\cite{zhang2025bountybench} in the cybersecurity landscape.

Such research highlights the critical nature of empirical evidence to understand which cybersecurity threats pose the greatest costs to an organisation, such that risk can be ascertained and threats can be prioritised against the cost of remediation.

Foley at al.~\cite{foley2022autonomous} highlight the asymmetric cost defenders have, given that an adversary only needs to succeed once whilst a defender must thwart each attack. The paper therefore proposes and evaluates reinforcement learning as an approach to level the playing field. However, such techniques and evaluations ultimately are dependent on how well the simulated attack scenarios match real-world attacker behaviour (i.e. the sim-to-real gap~\cite{jakobi95}). 

Zhang et al.~\cite{zhang2025bountybench} extend this knowledge by evaluating AI agents against real-world software vulnerabilities arising from bug bounty platforms, however a key gap is the underlying assumption that the financial reward of bug bounty programs matches the (financial) rewards of real-world attackers and the costs to an organisation if the vulnerability were exploited. As~\cite{freicase} notes, ``A larger number of vendors either lack the maturity, funding or incentive to invest more in secure software development''. Furthermore, a longitudinal analysis of bug bounty programs~\cite{walshe2022longitudinal} found that retaining offensive security professionals remains a common issue.

\subsection{Existing Datasets}
Harry et al~\cite{harry2018classifying} proposes a taxonomy for classifying cybersecurity events and provides a review of 3,355 events from news sources, filtered down to 2,431 after excluding events which did not have sufficient data. The authors found the top attack events were: exploitation of application server ($45.1\%$), ``message manipulation'' (i.e. defacement or hijacking a social media page, $17.8\%$), exploitation of end hosts ($17.3\%$) and external denial of service ($10.8\%$). However, this data is limited only to publicised events and specifically focuses on the effects of attacks instead of the tactics. Accordingly, this paper does not assist defenders seeking to identify which specific tactics are the most prevalent and how successful they are.

The European Union Agency for Cybersecurity (ENISA) has provided similar data for the transportation sector specifically~\cite{malatras2023enisa} for January 2021 to October 2022, which provides a useful insight into the attacks facing a sector that forms part of CNI. However, again this material is limited to threats rather than tactics. The types of attack listed included ransomware attacks ($38\%$), data related threats ($30\%$), malware ($17\%$), denial-of-service ($16\%$), phishing ($10\%$) and supply-chain attacks ($10\%$).

ENISA also published a 2024 report~\cite{enisa_threat_landscape_2024, enisa_state_of_cybersecurity_2024} covering July 2023 to June 2024 which provides a breakdown of approximately 11,079 incidents by threat type. The data is sourced from both open-source information and the agency's own cyber threat intelligence (CTI). The report indicates that $41.1\%$ relate to denial-of-service attacks, $25.79\%$ relate to ransomware, $19.01\%$ relate to ``data'', $6.06\%$ relate to social engineering threats and $5.19\%$ relate to malware, $0.98\%$ accounting for supply chain threats and $0.55\%$ related to FIMI (foreign information manipulations and interference). Web threats and zero-days are also mentioned as the remainder but no quantification is provided. However, the data is limited by being obtained from open-source information (though it's unclear to what extent versus ENISA's CTI) and does not provide a formal breakdown between attack type and initial access vector (with initial access vector data not specifically provided).

Various industry sources also seek to provide data on the initial access vectors for various cyberattacks. Verizon Data Breach Investigations Report (DBIR) 2025~\cite{verizon2025} provides information on initial access vectors sourced from both internal case files, third-party contributors and media reports, although the breakdown is not available. The data is sourced from 1 November 2023 to 31 October 2024. The breakdown for non-error, non-misuse breaches (n = 9,891) provided is that the initial access vectors in $22\%$ of instances was credential abuse, in $20\%$ of instances exploitation of vulnerabilities and in $16\%$ phishing attacks.

Mandiant, a subsidiary of Google, has also provided a similar report~\cite{mtrends2025} sourced from data from 1 Jan 2024 to 31 Dec 2024 where Mandiant was engaged on a consultancy basis. The number of cases is not described (though claimed to be based on 450,000+ hours of consultancy work), limited to those where consultancy work was utilised and $34\%$ of cases have an unknown initial access vector. Nonetheless, of those where the initial access vector was known $34\%$ were from exploits, $16\%$ stolen credentials, $14\%$ email phishing, $9\%$ web compromise, $8\%$ prior compromise, $7\%$ brute force, $5\%$ insider threat and $8\%$ ``other''.

CISA has provided a report~\cite{cisa2023} on attacks on CNI specifically for the Fiscal Year 2022. The data is limited to a pool of 121 RVAs (Risk and Vulnerability Assessments), which again is sourced through both open-source data and stakeholder assessment activities. The data provides a very different picture, with $54\%$ of initial access vectors being valid accounts, followed by spearphishing links as the second most successful compromise method. Additionally, the UK Government commissioned the research firm Ipsos to investigate experiences and impacts of ransomware attacks on individuals and organisations ~\cite{homeoffice_2025_ransomware_impacts}.

The UK National Cyber Security Centre (NCSC) annual review for 2024~\cite{ncsc_review} has shown that from September 2023 to August 2024, the agency has handled 430 incidents (an increase from 371 in the previous period), 347 data exfiltration incidents (increase from 327) and 89 ``highly significant and significant'' incidents (increase from 62). The report does not provide a breakdown of the types of attack or initial access vectors but states: ``Ransomware attacks continue to pose the most immediate and disruptive threat to our CNI, with some state-linked cyber groups now targeting the industrial control systems that infrastructure relies on''. No breakdown of initial access vector or further attack information is provided.

A further review published in 2025,~\cite{ncsc_review_2025} indicated the number of incidents handled in the following period remained largely the same (429, one less than the previous period), however there was a 50\% rise in ``highly significant'' attacks for the third consecutive year (18 incidents). There was still no breakdown of initial access vector, and ransomware was still highlighted as the major threat but no exact quantification was offered. A chart provided without numbers did highlight that $~10\%$ of cyber incidents categorised as ``highly-significant'' or ``significant'' were ransomware.

Finally, a report published by Waterfall Security~\cite{waterfall_security} has claimed that in 2024 ``87\% of identifiable attacks were ransomware''. The report only includes publicly reported incidents which have caused physical consequences in OT systems.

% 80\% of disruptive CNI incidents meant to be ransomware - source?

% THE CISA alerts and advisories are notable, although not essential to reference, - https://www.cisa.gov/news-events/cybersecurity-advisories?f%5B0%5D=advisory_type%3A94

Overall, there is limited empirical research demonstrating which attacks upon CNI achieved significant disruption, especially in academic literature.

\section{UK NIS Regulations Data}

\subsection{Methodology}
In the UK, Regulation 11 of the NIS Regulations 2018 places requirements on mandatory disclosure on ``operators of essential services'' to disclose to a DCA ``any incident which has a significant impact on the continuity of the essential service which that OES provides''. Regulation~12 goes on to place a similar requirement on operators of ``relevant digital service providers'' to notify the Information Commissioner ``about any incident having a substantial impact on the provision of any of the digital services''.

We seek to use the Freedom of Information Act 2000 (``FOIA'') and the Freedom of Information (Scotland) Act 2002 (``FOISA'') to evaluate the collection and recording of this data, and produce a dataset of initial access vectors of such cyber incidents. Whilst GCHQ is the ``single point of contact'' of which designated authorities must ultimately produce reports, it is not subject to FOIA and therefore we must contact each of the DCAs instead.

\begin{table}[h]
\caption{NIS Regulations Notifications}
\label{tab:nis-counts}
\centering
\scriptsize
\setlength{\tabcolsep}{3pt}
\renewcommand{\arraystretch}{1.1}
\begin{tabular}{@{}lrr@{}}
\hline
\textbf{DCA} & \textbf{Total Count} & \textbf{Cybersecurity Count} \\
\hline
Water (Scotland) & 0 & 0 \\
Health (Scotland) & 34 & 7 \\
Civil Aviation & 1 & 0 \\
Water (England \& Wales) & 9 (+ 5 non-NIS) & 3 (+ 5 non-NIS) \\
Energy (Scotland, England \& Wales) & 0 & 0 \\
Health (England \& Wales) & 6 (+ 48 non-NIS) & 5 (unknown non-NIS) \\
Digital Infrastructure & 4 & 2 \\
Transportation (England \& Wales) & 11 & 3 \\
Relevant Digital Service Providers & 5 (+ 6 non-NIS) & 2 \\
Northern Ireland Government & 33 & 8 \\
\hline
\end{tabular}
\end{table}

\begin{table}[h]
\caption{NIS Notifications with Cybersecurity Incident Specifics}
\label{tab:nis-incidents}
\centering
\scriptsize
\setlength{\tabcolsep}{3pt}
\renewcommand{\arraystretch}{1.1}
\begin{tabular}{@{}ll@{}}
\hline
\textbf{Incident} & \textbf{DCA} \\
\hline
5 SMS-Teknik Data Breach Incidents~\cite{sms_teknik_2024} & Health (Scotland) \\
WordPress Plugin Infection & Health (Scotland) \\
Kerberoasting & Health (Scotland) \\
5 Ransomware Incidents & Health (England \& Wales) \\
2 Exploitation of Vulnerabilities Incidents & Relevant Digital Service Providers \\
DDoS (Unknown Type) & Digital Infrastructure \\
DDoS (DNS Query Flood) & Digital Infrastructure \\
\hline
\end{tabular}
\end{table}

\subsection{Results}

The requests were made from the 25 to 30 September 2025. Responses began to be received from the 30 September 2025 and were resolved by the 20 November 2025. A summary of the data is provided in Table~\ref{tab:nis-counts} and where cybersecurity incident specifics have been provided, they are summarised in Table~\ref{tab:nis-incidents}. Details of the responses for each of the DCAs contacted by means of FOI request are provided below. \newline

\noindent\textbf{Water (Scotland):}
The Drinking Water Quality Regulator (DWQR) for Scotland stated they ``received no NIS reports in 2024''. An early FOI request published by the Scottish Government~\cite{scot_gov_foia} confirmed the DWQR had received no reports but the Scottish Government had received 20 reports under the NIS Regulations generally during 2024 (11 of the 20 did not have the incident report submitted to the Government within 72 hours of the incident occurring).\\

\noindent\textbf{Health (Scotland):}
We received a specific response dated the 7th October 2025 from the Scottish Government's Directorate for Population Health related to reports received by Health Boards in Scotland. The Scottish Government confirmed that at the latest data, there were a total of 34 incidents reported and 7 were cybersecurity related. 5 of the healthcare cybersecurity incidents related to a data breach related to a third party supplier, SMS-Teknik~\cite{sms_teknik_2024}. A further incident ``involved an infected WordPress plugin on a corporate site, resulting in unauthorised links to adult content and a cryptocurrency miner''. Finally, one incident categorised under ``cyber'' was described as follows: ``1 incident involved suspicious activity, including Kerberoasting, a technique that targets the Kerberos authentication protocol led to the immediate shutdown of servers''.\\

\noindent\textbf{Civil Aviation (UK):}
The Civil Aviation Authority confirmed they had received one notification for the sub-sector they were the DCA during 2024, however stated it did not relate to a cybersecurity incident.\\

\noindent\textbf{Water (England \& Wales):}
The Drinking Water Inspectorate said they received 9 reports within the scope of the NIS Regulations (8 in England and 1 in Wales) and 5 reports which were below the threshold of the NIS Regulations purely relating to England. 3 of those within the scope of the NIS Regulations and all 5 below threshold related to cybersecurity incidents. The DCA refused to share high-level categorisation of the attack types and initial access vectors citing national security grounds, for which an internal review to their decision was subsequently refused which we did not appeal to the Commissioner.\\

\noindent\textbf{Energy (Scotland, England \& Wales):} The Department for Energy Security and Net Zero stated: ``we can confirm that in the 2024 calendar year, there were no reported NIS Incident in the energy sector''.\\

\noindent\textbf{Health (England \& Wales):} 
The Department of Health and Social Care stated: ``54 reports were made to DHSC in 2024. These related to a mix of cybersecurity and other broader resilience factors impacting network and information systems (as specified in the scope of the NIS Regulations 2018)''. They went on to say ``Of the six incidents that met the thresholds, five related to ransomware and one to a power outage''. They did not hold data on the initial access vectors of the cyberattacks.\\

\noindent\textbf{Digital Infrastructure (UK):} 
The Information Commissioner receives reports from Relevant Digital Service Providers, where a service was unavailable for $> 750{,}000$ user-hours, there was data integrity/confidentiality/authenticity loss affecting $> 15{,}000$ UK users, the incident created a risk to public safety, public security, or loss of life or the incident caused material damage to at least one UK user exceeding \textsterling~850,000~\cite{ico_reporting}. They responded: ``Of the 5 reports considered to be NIS incidents, 2 were assessed as cyber incidents. The type of attack was hacking, and the access vector was exploitation of vulnerabilities''.

In relation to Digital Infrastructure (DI), Ofcom initially extended the 20 day deadline for a response to 40 days in order to consider whether an exemption relating to law enforcement applied to the request and whether the public interest factors were in favour of releasing the information. We responded with submissions on the legal basis for disclosure, technical arguments in favour of the release and comparators showing other organisations had disclosed the information. Ofcom then disclosed that: ``One mandatory and three voluntary incident reports relating to the DI subsector were submitted to Ofcom in 2024''. Two of these were disclosed to be related to cybersecurity, but a breakdown was not supplied.

More interestingly, however, Ofcom provided the following information in relation to attack type: ``Both of the two cybersecurity attacks reported to Ofcom during 2024 were DDoS (Distributed denial-of-service) attacks''. In relation to initial access vectors they said: ``Only one of the DDoS attack reports provided an initial attack vector which was a DNS query flood''.\\

\noindent\textbf{Transportation (England \& Wales):} 
The UK Government's Department for Transport reported there had been 1 disclosure related to maritime transport, 3 related to rail and 7 related to road. All 3 NIS incidents related to rail were cybersecurity incidents. They refused to disclose type of attack and initial attack vector data.\\

\noindent\textbf{Government (Northern Ireland):}
Whilst the Northern Ireland Department of Finance (DoF) initially refused to disclose any information, following negotiation we were able to obtain a disclosure that on a province-wide basis there had been 33 NIS incidents reported in 2024, 8 were cybersecurity incidents and a statement that ``None of the reported incidents were as a direct cyberattack on the organisation''.\\

\noindent\textbf{Government (Wales):}
The Welsh Government stated: ``The information requested is not held''.\\

\noindent\textbf{Maritime \& Coast (UK):}
We additionally contacted the Maritime \& Coastguard Agency, which whilst not a DCA, it is an executive agency of the Department for Transport (which is a DCA), who stated: ``We can confirm that the Maritime and Coastguard Agency (MCA) does not hold the information detailed in your request which we have taken to mean reports made to the MCA under The Network and Information Systems Regulations 2018. To conclude that the information is not held, we consulted with our IT and Information Security teams. There have been no reports made to the MCA during the period requested''. Similarly, National Highways confirmed they did not hold the information, as it was not a DCA.

\subsection{Analysis}
We note there appears to be differences in the thresholds as to what different DCAs consider to be ``any incident which has a significant impact on the continuity of the essential service which that OES provides''~\cite{ico_reporting, dhsc_reporting}. Having different thresholds for different elements of CNI can make sense to evaluate disruption, however this can mean that comparing data across industries for cybersecurity purposes is not on a like-for-like basis.

We also note that there seems to be greater attack-type detail on the data obtained related to healthcare in Great Britain, perhaps due to the public provision of healthcare services. However, this highlights the critical issue that ransomware poses on healthcare. Healthcare amounted to $55\%$ of all cybersecurity incidents in our dataset (excluding Northern Ireland). The healthcare data from England \& Wales highlights that all incidents were ransomware attacks. In Scotland, the 5 reports related to the SMS-Teknik data breach~\cite{sms_teknik_2024} involved a ransom note being left by the attackers and a further incident involving Kerberoasting could be used as part of a ransomware attack and the final incident involving defacement of a WordPress website ``unauthorised links to adult content and a cryptocurrency miner'' clearly also can contain a profit motive. Therefore we estimate that $83\%-92\%$ of cyberattacks in Great Britain during 2024 which had a significant impact on the continuity healthcare services were as a result of ransomware, and $92\%-100\%$ contained a profit motive.

One limitation is that we have only queried data for 2024. Both FOIA and FOISA contain proportionality limits in relation to limiting overly costly and ``vexatious'' requests. Additionally, requesting more recent data could have caused more data to be subject to refusals on national security grounds. Using the methodology we have developed here, future work may wish to attempt to go deeper into the past with greater data or seek to obtain more recent data as time progresses.

Our data also highlights how a single incident could be reportable multiple times. This can also be seen in the Scottish healthcare data where a single SMS-Teknik breach affected multiple Health Boards. In principle, this could also lead to one incident being reported across multiple DCAs (e.g. if English health authorities had also used the same vendor). The unit of measurement is therefore focussed on impact to individual operators of essential services. 

Our dataset found there were 30 cybersecurity reports within the scope of the NIS Regulations. We contrast this with the 2024 figure (albeit from reporting window, September 2023 to August 2024) from the NCSC indicating there were 89 ``highly significant and significant'' incidents. If the distribution were to remain uniform for the rest of the window, this indicates that potentially around two-thirds of ``highly significant and significant'' incidents reported to the NCSC may not currently be captured under the NIS Regulations. The true extent of under-reporting is likely to be greater given the following reporting year, the number of nationally significant incidents NCSC reported increased from 89 to 204.

It is unclear what the cause of such under-reporting is. One cause of this may be divergence between the classifications used by NCSC~\cite{ncsc_categorisation} and those used by DCAs, however there is a high bar for both types of incident classification. Another potential cause is that the NCSC has a broader focus than just what is classified as Critical National Infrastructure under the NIS Regulations, covering other large organisation and government bodies/authorities. Additionally, NCSC can source data from other sources than just mandatory disclosure regimes.

Whilst this makes a powerful argument for wider mandatory reporting of ransomware data that the UK Government consulted on~\cite{government_ransomware_consultation}, our research has highlighted that such work must ensure reporting is made available to researchers. Whilst the NCSC has published annual reviews~\cite{ncsc_review, ncsc_review_2025}, the granularity of data would not have been available had we not been able to workaround the fact that the NCSC is exempt from FOI laws by making requests to DCAs instead. It is therefore essential that such mandatory reporting regimes contain mechanisms for academic researchers and policymakers to gain access to such data, such that cybersecurity research and public policy can be as best informed as possible.

\section{CISA Alerts \& Advisories}
\subsection{Methodology}
The US CISA, alongside other agencies (both domestically and internationally) publishes cybersecurity alerts and advisories~\cite{cisa_advisories} in order for practitioners to mitigate cyberthreats. The data also potentially provides a source of data to understand the causes of cyberattacks. Whilst the dataset is more curated, it also provides a specific picture of initial access vectors by attack type which is missing in much of the data we have collected from UK DCAs.

We seek to review the reports published during 2024 (although not necessarily concerning events in 2024), to understand and classify attacks according to the type of attack using the MITRE ATT\&CK~\cite{mitre_attack} and VERIS~\cite{veris_framework} frameworks.

\subsection{Results}
There were 17 reports published during 2024, we find two which reference simulated attacks and one that discusses trends:
\begin{itemize}
    \item AA24-326A – \textit{CISA Red Team Assessment (critical infrastructure)}.
    \item AA24-193A – \textit{CISA SILENTSHIELD Red Team (FCEB)}.
    \item AA24-317A – \textit{2023 Top Routinely Exploited Vulnerabilities}.
\end{itemize}

The other 14 reference real-world attacks. The attacks and the reported initial access vector are provided in Table~\ref{tab:attacks-compact} and the count for each according using the MITRE ATT\&CK Framework are provided in Table~\ref{tab:ia-counts}. In relation to the ATT\&CK access vector counts, one incident can involve multiple techniques, so the total count can be greater than 14; additionally, brute-force/password spraying and MFA-prompt fatigue were mapped to \emph{Valid Accounts} and/or \emph{External Remote Services} for initial access accounting.

\begin{table}[t]
\caption{CISA Reported Attacks \& Initial Access Vectors}
\label{tab:attacks-compact}
\centering
\scriptsize
\setlength{\tabcolsep}{3pt}
\renewcommand{\arraystretch}{1.1}
\begin{tabular}{@{}lll@{}}
\hline
\textbf{Report ID} & \textbf{Type of Attack} & \textbf{Initial Access Vector} \\
\hline
AA24-016A & Malware / intrusion & Exploit vuln (web) \\
AA24-038A & Espionage & Exploit vuln (edge) \\
AA24-046A & Credential misuse (insider) & Cred abuse (ex-employee) \\
AA24-060A & Ransomware (Phobos) & Cred abuse (RDP) \\
AA24-060B & Intrusion & Exploit vuln (SSL-VPN) \\
AA24-109A & Ransomware (Akira) & Cred abuse (VPN/RDP) \\
AA24-131A & Ransomware (Black Basta) & Phishing \\
AA24-207A & Espionage & Exploit vuln (web) \\
AA24-241A & Ransomware (IAB/enablers) & Exploit vuln (edge) \\
AA24-242A & Ransomware (RansomHub) & Phishing \\
AA24-249A & Espionage & Exploit vuln (web/edge) \\
AA24-290A & Credential abuse & Cred abuse (spray/MFA) \\
SVR-Cloud (NCSC) & Espionage (cloud) & Cred abuse (spray/MFA) \\
APT40 Advisory (ASD) & Espionage & Exploit vuln (web) \\
\hline
\vspace{0ex}
\end{tabular}
\footnotesize Vectors aligned to VERIS categories: \emph{Exploit vuln} (Hacking$\rightarrow$Exploit vuln), \emph{Cred abuse} (Hacking$\rightarrow$Use of stolen creds), \emph{Phishing} (Social$\rightarrow$Phishing).
\end{table}

\begin{table}[t]
\caption{ATT\&CK Initial Access Techniques (Count)}
\label{tab:ia-counts}
\centering
\scriptsize
\setlength{\tabcolsep}{3pt}
\renewcommand{\arraystretch}{1.1}
\begin{tabular}{@{}lrr@{}}
\hline
\textbf{ATT\&CK Initial Access} & \textbf{Count} & \textbf{Share} \\
\hline
T1190 Exploit Public-Facing Application & 10 & 71\% \\
T1133 External Remote Services & 7 & 50\% \\
T1078 Valid Accounts & 6 & 43\% \\
T1566 Phishing & 4 & 29\% \\
\hline
\end{tabular}
\end{table}

\subsection{Analysis}
This is a curated dataset which is based on advisories that CISA (or Five Eyes allies, namely the UK's GCHQ and the Australian Signals Directorate) have chosen to publish so may not represent comprehensively the real-world attack landscape. This is demonstrated by the data from the UK's NCSC~\cite{ncsc_review} showing they handled 430 incidents in one reporting period alone (not to mention 1957 total tip offs).

However, it is of note that when considering the primary initial access vector, 50\% related to human factors (phishing or credential abuse) and the other 50\% related to some form of exploit vulnerability (i.e. web or edge network). Additionally, the focus areas the agency have chosen to address are of interest: 5 ($35.7\%$) relate to ransomware and a further 5 ($35.7\%$) relate to espionage whilst a further $4$ relate to other attacks (e.g. insider credential misuse, $28.6\%$).

Of 5 attacks which related to ransomware, 4 initial access vectors related to credential abuse or phishing, whilst only one related to the exploitation of vulnerabilities - indicating that authentication security can be a key concern in relation to ransomware.

Nevertheless, the CISA advice remains focussed on patching external access vectors, hardening identity systems and monitoring cloud environments.

\section{Cyber Security and Resilience Bill}
We find that 29\% (i.e. 30 of 103) NIS reports in 2024 related to cybersecurity incidents, despite the purpose of the NIS Regulations being targeted at network and information systems more broadly. However, this research has identified deficiencies in the existing regulatory environment which requires consideration for further legislation. The NCSC Annual Review 2025~\cite{ncsc_review_2025} also makes the following points:

\begin{itemize}
  \item ``Our sharing communities need to be deeper, faster and more actionable, sharing data and insight at speed, driving quicker evidence-based decision making''.
  \item ``We are committed to automating that evidence base and making it transparent, to invite and encourage challenge and review''.
  \item ``Technology is currently opaque, and that leads to adverse cybersecurity outcomes, which favour malfeasant threat actors''.
\end{itemize}

The need for focussed regulation on cybersecurity has already seen policymaking progress in the UK with the proposals for a Cyber Security and Resilience Bill~\cite{cybersecurity_policy_statement}, mirroring similar efforts of the NIS2 Directive in the EU and a recent consultation of mandatory reporting of ransomware~\cite{government_ransomware_consultation}.

On the 12th November 2025, the UK Government introduced the Cyber Security and Resilience (Network and Information Systems) Bill to Parliament~\cite{uk_gov_bill}. The Bill essentially has the effect of expanding the coverage of the NIS Regulations to bring into scope data infrastructure (i.e. data centres), managed service providers and critical suppliers. The Bill requires initial notification within 24 hours and then full notification within 72 hours (though remains silent on a full root-cause investigation). Additionally, there are requirements to send notifications to GCHQ at the same time as to DCAs. Furthermore, the definition of ``incident'' in the NIS regulations is sought to be broadened from ``any event having an actual adverse effect on the security of network and information systems'' to ``any event having, or capable of having, an adverse effect on the operation or security of network and information systems''.

These measures therefore address many of the limitations identified in this paper. However, whilst the methodology outlined in this paper remains effective if the Bill is enacted as drafted, academic researcher access to the data remains an open question. Working around GCHQ's exemption from FOIA and FOISA by querying each DCA (and the Information Commissioner) individually is time-consuming and requires significant education of those involved in the disclosure process: we made 22 FOI requests in total, with substantial follow-up to educate decision-makers factors for disclosure.

For example, the Northern Irish DoF initially refused to provide any data at all including high-level numeric data of either NIS reports or the counts of those specific to cybersecurity. Through both education and negotiation we were able to negotiate province-wide count data.

However, if a single source of truth was available, those responsible for disclosure would be able to provide aggregated data for the fields desired (i.e. a breakdown of all attacks by sector, attack type or initial access vector), which could mitigate issues whereby small datasets could be used to target particular sectors in a particular province.

Both the public reports from NCSC and the data we have analysed from CISA highlight that data from intelligence services is currently curated, limiting use for security researchers. Accordingly, we highlight this gap to demonstrate that mandatory publication of research data either in new legislation or agreements for voluntary reporting with such organisations. This would aid in the NCSC's goal of transparency of the evidence with their sharing communities.

\section{Key Lessons Learned \& Recommendations}
Mandatory reporting of cybersecurity is not unique to the UK. Indeed, there have been developments in the EU, Australia, Singapore, South Korea, India, China and the USA~\cite{cybersecurity_policy_statement, seng2023cybersecurity, seng2024cybersecurity, cisa_legislation}. Our methodology of combining mandatory disclosure laws with freedom of information laws may benefit researchers globally in collecting their own data.

Highlighting how global policymakers should ensure reporting results in actionable threat intelligence data, our research found that DHSC in England did not hold initial access data on the ransomware attacks faced.

Finally, CISA advisories on ransomware attacks concerned edge or authentication initial access; Scottish healthcare similarly saw third-party data breaches, authentication system attacks and web vulnerabilities exploited - highlighting the importance of authentication and edge security for practitioners, even where the headline attack type is ransomware.

\section{Conclusion}
Firstly, as policymakers consider the impact of new cybersecurity reporting requirements, in relation to \textbf{RQ2}, we identify deficiencies in the existing NIS Regulations. We find that there were significantly fewer NIS notifications than incidents classified as ``highly significant and significant'' by the NCSC in their 2024 reporting year. We also note that detailed initial access vector data was either missing or disclosure was refused. This highlights a key limitation in existing reporting regimes in how they both fail to capture a complete picture in relation to cybersecurity, and data is limited for academic research and policymakers.

Secondly, turning to \textbf{RQ3}, even amongst the curated CISA dataset, we note that when considering the primary initial access vector, 50\% related to human factors (phishing or credential abuse) and the other 50\% related to some form of exploit vulnerability (i.e. web or edge network). This highlights the need for cybersecurity research to focus on edge network security and human factors.

Finally, we note that profit motive seems increasingly to drive attacks which impact the continuity of critical services. Whilst $36\%$ of CISA reported attacks concerned espionage, based on the NIS data, we estimate that $83\%-92\%$ of cyberattacks in Great Britain during 2024 which had a significant impact on the continuity healthcare services were as a result of ransomware, and $92\%-100\%$ of attacks contained a profit motive. This aligns with Waterfall Security's claim~\cite{waterfall_security} that in 2024 ``87\% of identifiable attacks were ransomware'' among publicly reported incidents causing physical consequences in OT systems.

With regards to \textbf{RQ1}, whilst we demonstrate that mandatory-disclosure regimes can be combined with FOI laws to assist researchers in collecting data - our research highlights the need for providing a single source of data for mandatory cybersecurity regimes, such that the data can be aggregated together in a way which is useful for policymakers and cybersecurity researchers. In the words of the NCSC~\cite{ncsc_review_2025}: ``Technology is currently opaque, and that leads to adverse cybersecurity outcomes, which favour malfeasant threat actors''. A single source offers the ability to aggregate national data, providing breakdowns by sectors and attack vectors, whilst avoiding the risk that data is de-anonymised.

As the UK Government implements greater mandatory reporting regimes, further work is needed to obtain and analyse such data, providing better signals to researchers and policymakers on which areas to target. Future work may also apply these methodologies in other jurisdictions, such as the EU and United States, using their respective FOI legislation.

\bibliographystyle{plain} % We choose the "plain" reference style
\bibliography{bibliography}

\end{document}